\documentclass{article}

\usepackage{microtype}
\usepackage{graphicx}
\usepackage{subfigure}
\usepackage{siunitx}
\usepackage{booktabs} 

\usepackage{hyperref}

\usepackage[accepted]{icml2020}

\DeclareSIUnit[number-unit-product = {\,}]\molar{M}

\icmltitlerunning{Attribution Methods Reveal Flaws in Fingerprint-Based Virtual Screening}

\begin{document}

\twocolumn[
\icmltitle{Attribution Methods Reveal Flaws in Fingerprint-Based Virtual Screening}



\icmlsetsymbol{equal}{*}

\begin{icmlauthorlist}
\icmlauthor{Vikram Sundar}{goog}
\icmlauthor{Lucy Colwell}{goog,cam}
\end{icmlauthorlist}

\icmlaffiliation{goog}{Google Research, Mountain View, CA, USA}
\icmlaffiliation{cam}{Department of Chemistry, University of Cambridge, Cambridge, Cambridgeshire, UK}

\icmlcorrespondingauthor{Lucy Colwell}{ljc37@cam.ac.uk}

\icmlkeywords{Protein-Ligand Binding, Attribution, Interpretability, Virtual Screening}

\vskip 0.3in
]



\printAffiliationsAndNotice{}  

\begin{abstract}
    Fingerprint-based models for protein-ligand binding have demonstrated outstanding success on benchmark datasets; however, these models may not learn the correct binding rules. To assess this concern, we use \textit{in silico} datasets with known binding rules to develop a general framework for evaluating model attribution. This framework identifies fragments that a model considers necessary to achieve a particular score, sidestepping the need for a model to be differentiable. Our results confirm that high-performing models may not learn the correct binding rule, and suggest concrete steps that can remedy this situation. We show that adding fragment-matched inactive molecules (decoys) to the data reduces attribution false negatives, while attribution false positives largely arise from the background correlation structure of molecular data. Normalizing for these background correlations helps to reveal the true binding logic. Our work highlights the danger of trusting attributions from high-performing models and suggests that a closer examination of fingerprint correlation structure and better decoy selection may help reduce misattributions.
\end{abstract}

\section{Introduction}

Identifying ligands that bind tightly to a given protein target is a crucial first step in drug discovery. Experimental methods such as high-throughput screening are time-consuming, and costly, while physics-based methods are computationally expensive and can be inaccurate \cite{MacConnell2017, Schneider2017, Grinter2014, Chen2015}. The emergence of large datasets enables data-driven approaches to be applied to this problem. In recent years, a variety of ML-based approaches have been developed to identify active ligands for a protein target given screening data \cite{Gawehn2016, Colwell2018, Ripphausen2011}. These approaches report outstanding {\it in silico} success on benchmark datasets \cite{Ragoza2017, Ramsundar2017, Gomes2017}. 

However, recent studies analyzing how neural network models attribute their results suggest that even high-performing models often do not learn the correct rule \cite{McCloskey2019}. Fingerprint-based models also share these issues; \citet{Sheridan2019} observed that high-performing models of the same dataset attribute binding activity to different atoms within a molecule. Attribution of virtual screening models is particularly important because accurate identification of pharmacophores would enable medicinal chemists to improve potential drug candidates. 

In this paper, we evaluate attributions from fingerprint-based virtual screening models, complementing a previous analysis of graph convolutional models \cite{McCloskey2019}. We propose a general framework for evaluating model attributions that does not require gradients, and use \textit{in silico} datasets to evaluate a number of standard models. In agreement with previous work \cite{Sheridan2019}, our results establish that high-performing models may not learn the correct binding rule. Going beyond previous work, we analyze properties of the data that lead to misattributions, and provide insight into how this can be mitigated. Our analysis reveals that attribution results can be improved by (i) adding fragment-matched decoys, and (ii) accounting for spurious correlations in the data that originate from both the nature of small molecule structures and the definition of fingerprint descriptors.

\section{Methods}

\subsection{Datasets and Models}

To measure attribution we construct a number of \textit{in silico} datasets, where each dataset has a specified binding logic that requires $3$ randomly selected fragments to be present for a molecule to be active. We identified $600$ active and $600$ inactive ligands from the ZINC12 database \cite{Irwin2012} using each binding logic to generate each dataset. Dataset 0 required a benzene, alkyne, and amino group to be considered active; dataset 1 an alkyne, benzene, and hydroxyl; dataset 2 a fluorine, alkene, and benzene; and dataset 3 a benzene, ether, and amino group. Our binding logics mimic known pharmacaphores for real proteins; for example, most binders to soluble epoxide hydrolase have an amide and a urea group \cite{Waltenberger2016}.

We used ECFP6 fingerprints with 2048 bits \cite{Rogers2010} to featurize molecules. We tested Naive Bayes, Logistic Regression, and Random Forest models, all implemented using scikit-learn \cite{Pedregosa2011}. Naive Bayes used no prior; Logistic Regression used $C = 1$; Random Forest used $100$ trees and a maximum depth of $25$. Model AUCs (Area under the Receiver Operator Characteristic curve) were computed on a randomly held-out test set comprising $20\%$ of the data, with an even split of actives and inactives. All train/test splits were repeated $50$ times to measure the AUC variation.

\subsection{Evaluating Attribution}

For attribution analysis we retrained the models on all the data, without any hold-out sets. Since some models were not differentiable, we developed an attribution method that works for any fingerprint-based model. Our method relies on SIS (sufficient input subsets), which identifies sufficient subsets of the input features to reach a particular score threshold for the given model \cite{Carter2018}. Our null feature mask was the vector of all $0$s. If our model predicted a binding probability of $x$ for a given molecule, the threshold used for SIS attribution was $\frac{\lceil 100 (x-0.01) \rceil}{100}$.

The next step in our attribution procedure is to translate the significant features (fingerprint bits) back to fragments. For each molecule, we used rdkit to indicate the relevant fragment or fragments for each bit \cite{Landrum2006}. For a pre-trained model $m$ and a specific molecule, we generate an attribution vector $\mathbf{v}_m$ of dimensionality the number of atoms in the molecule. Each element of $\mathbf{v}_m$ corresponds to an atom in the molecule and is the number of significant fragments containing that atom. When a single bit corresponds to multiple fragments, all are considered significant. The ground truth attribution vector $\mathbf{v}_R$ for each molecule has $\mathbf{v}_R = 1$ for any atom belonging to fragment in the binding logic and $\mathbf{v}_R = 0$ otherwise. 

We used $\mathbf{v}_m$ and $\mathbf{v}_R$, to compute four metrics of attribution accuracy. First, for a given molecule the cosine similarity $\frac{\mathbf{v}_m \cdot \mathbf{v}_R}{||\mathbf{v}_m|| ||\mathbf{v}_R||}$ provides an attribution score for the model $m$. This score is normalized by the attribution score for the same model trained on randomly labeled data to avoid data structure effects \cite{Adebayo2018}. The attribution false positive score is the proportion of atoms that the model erroneously considers relevant. The attribution false negative score is the proportion of relevant atoms that the attribution misses. Finally, the comparative attribution score between two models $m_1$ and $m_2$ is $\frac{\mathbf{v}_{m_1} \cdot \mathbf{v}_{m_2}}{||\mathbf{v}_{m_1}|| ||\mathbf{v}_{m_2}||}$. We computed attribution scores for $50$ randomly selected molecules, each predicted to bind with probability at least $0.95$, to compute errors in attribution scores.

\section{Attribution Results}

\begin{figure}[!ht]
    \centering
    \includegraphics[width=0.45\textwidth]{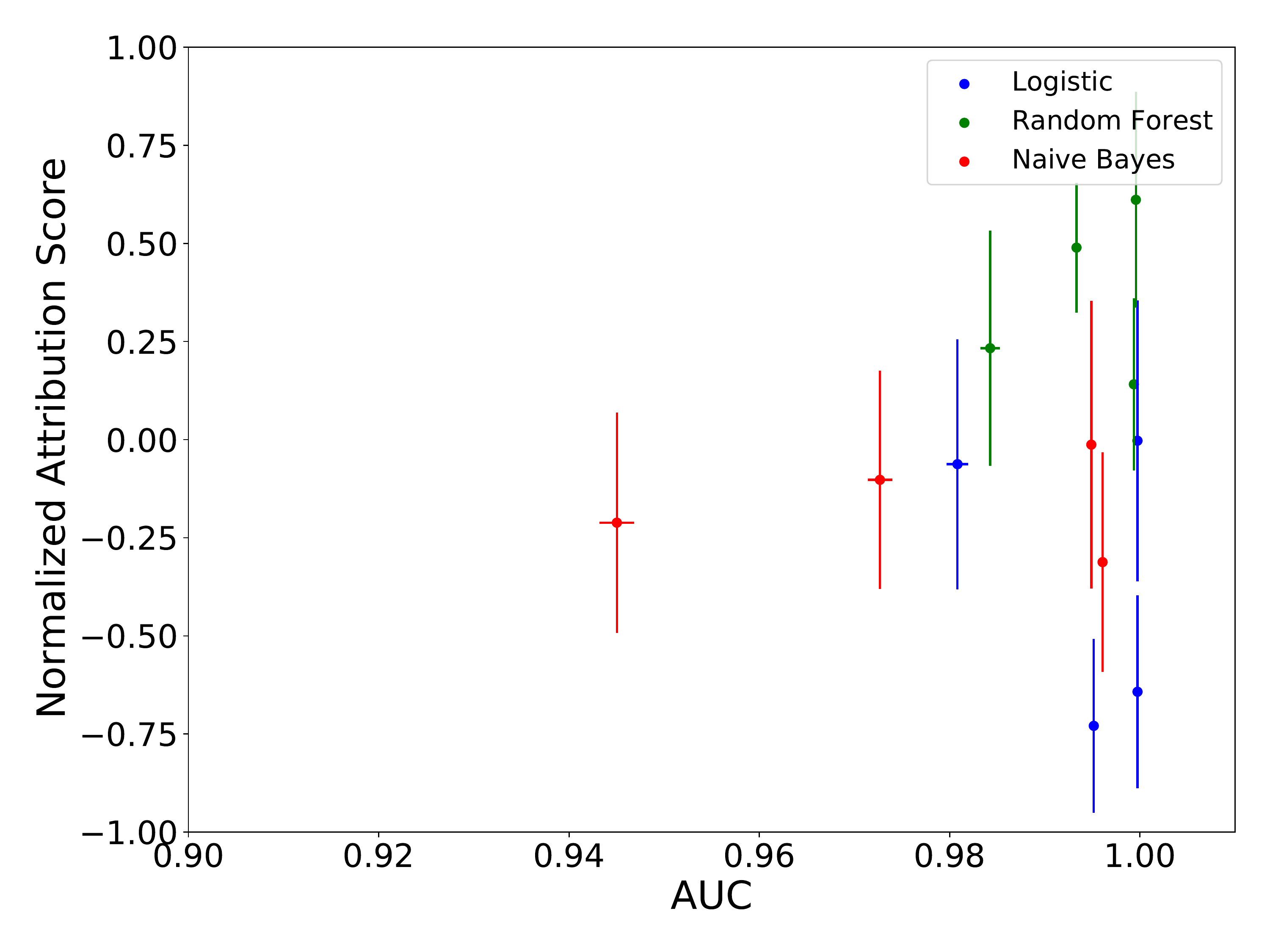}
    \caption{Comparison of attribution scores to AUC on \textit{in silico} datasets. The attribution score is a measure of how close the model's attribution is to the real rule for a particular molecule. Even high-performing models have poor attribution scores, suggesting that they are learning the wrong rule.}
    \label{fig:ToyDatasetAnalysis}
\end{figure}

The models, especially logistic and random forest, perform outstandingly well on the \textit{in silico} datasets; Figure \ref{fig:ToyDatasetAnalysis} shows that many models achieve very high AUCs. However, the normalized attribution scores in Figure \ref{fig:ToyDatasetAnalysis} are much lower. Even high-performing models tend to generate poor attribution scores with large variance between molecules, suggesting that the models are learning something very different from the ground truth logic.

\begin{figure}[!ht]
    \centering
    \includegraphics[width=0.45\textwidth]{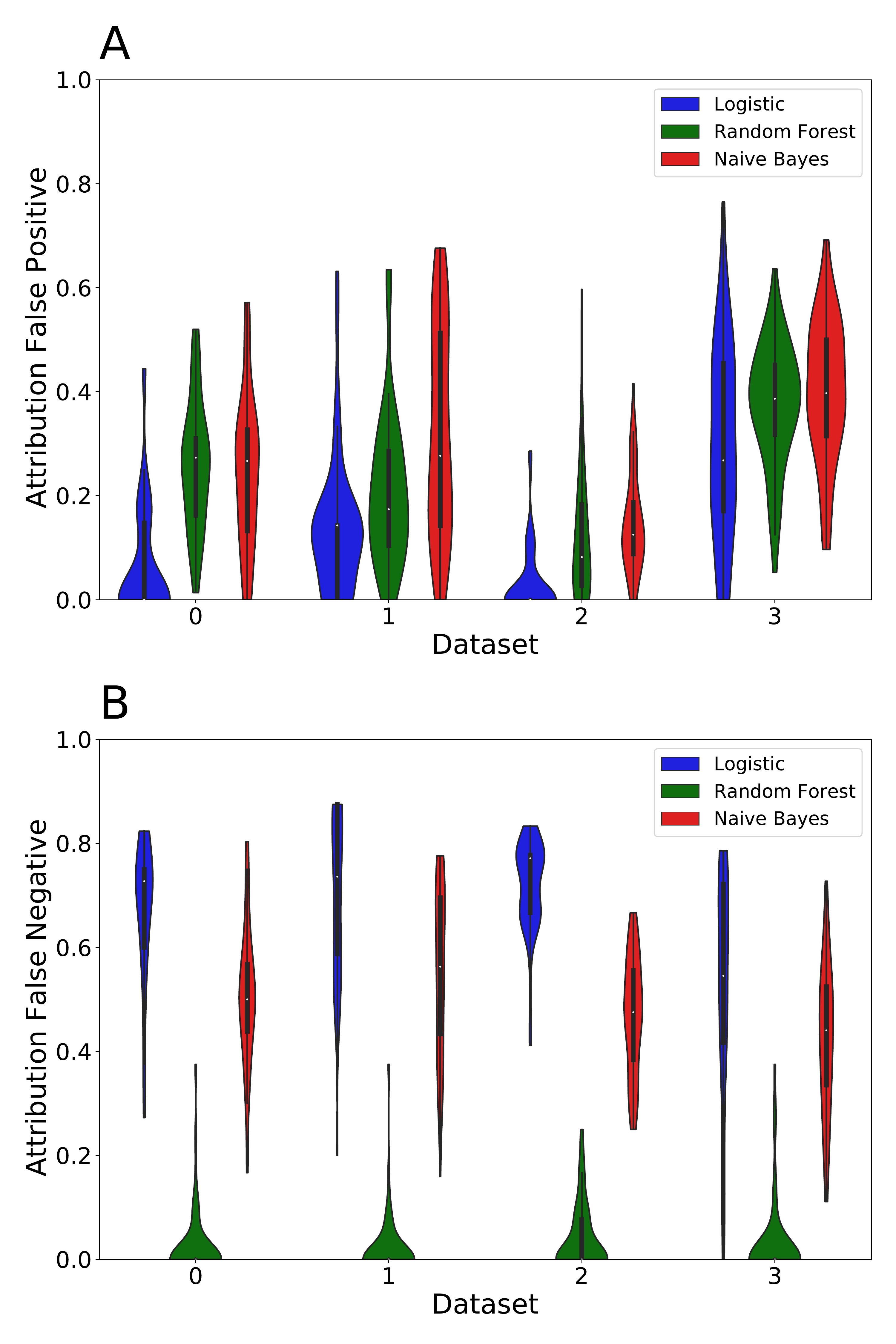}
    \caption{(a) Attribution False Positive and (b) False Negatives. We observe higher rates of attribution false negatives for the logistic regression models. Random forest models have higher rates of false positives.}
    \label{fig:ToyDatasetFPFN}
\end{figure}

\begin{figure}[!ht]
    \centering
    \includegraphics[width=0.5\textwidth]{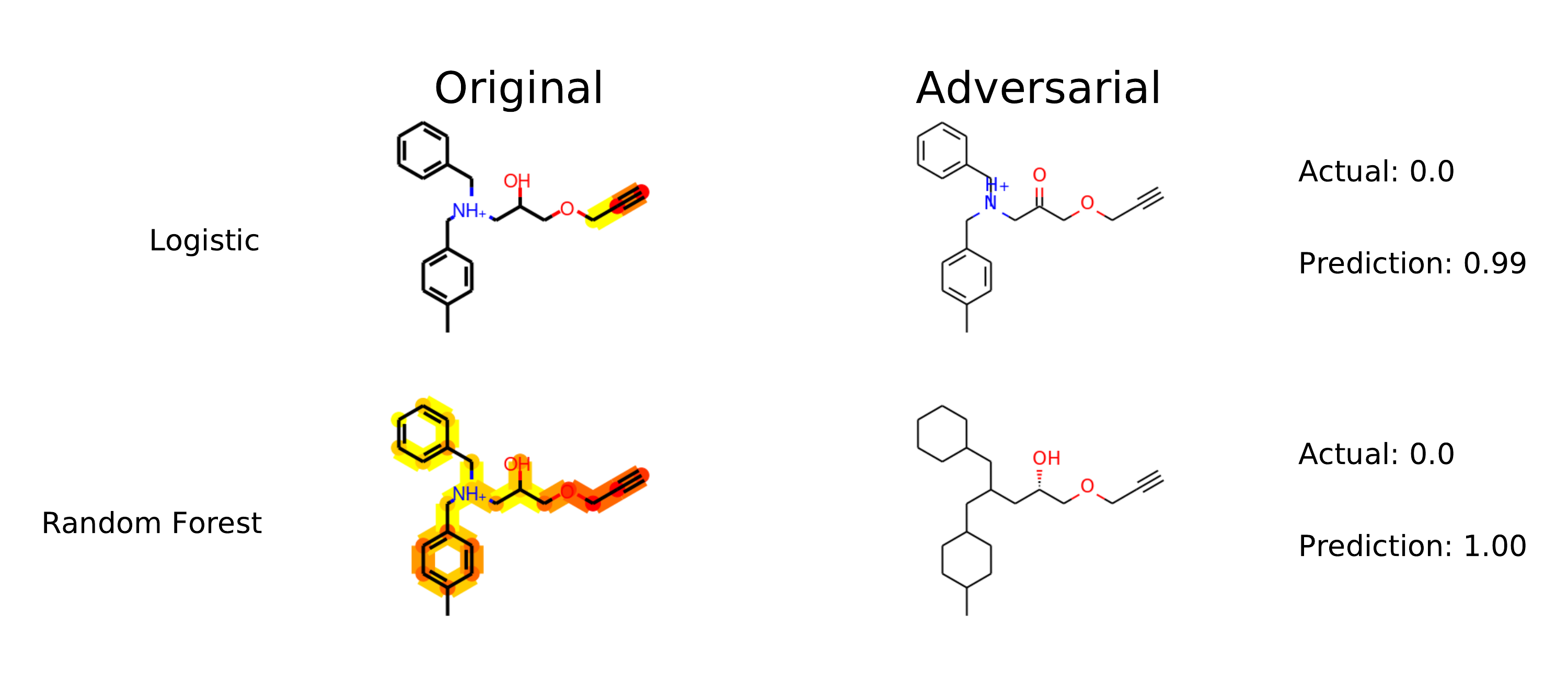}
    \caption{Attribution Images and Adversarial Examples for Dataset 1, with binding rule benzene, alkyne, and hydroxyl. The left molecule is the attribution image for the specified model. Red has highest attribution; white has lowest. The right image is an adversarial example constructed using the observed misattributions. The success of the adversarial examples indicates that our attribution method correctly identifies flaws in model performance.}
    \label{fig:ToyDataThreeElemAdversarial}
\end{figure}

To better understand this data, Figure \ref{fig:ToyDatasetFPFN} splits the attribution score into false positives and false negatives, while Figure \ref{fig:ToyDataThreeElemAdversarial} shows some attribution images for dataset 1. Both suggest that logistic regression is more susceptible to false negatives, while random forest is more susceptible to false positives. These results call into question the validity of using these models to physically interpret predictions or to garner medicinal chemistry insight. In order to further establish that our attribution results provide insight into model performance, we used knowledge of the misattributed atoms to manually generate adversarial examples; a sample are shown in Figure \ref{fig:ToyDataThreeElemAdversarial}. This shows that erroneous attributions correspond to weaknesses in the trained models.

\subsection{Misattribution: Fragment-Matched Decoys}

To understand attribution false negatives, we compute the Pearson correlation between the presence of every feature and the binding activity of each molecule and examine those features most correlated with activity. We observe that benzene does not appear among the top $20$ features for dataset 1 (data not shown), despite the fact that it is required in the binding logic. This is despite the fact that the top $20$ contains features that are not present in the binding logic. This surprising observation likely explains why models fail to place high weight on benzene, as seen in Figure \ref{fig:ToyDataThreeElemAdversarial}. To address this issue, we add fragment-matched decoys: inactive molecules that have some, but not all, of the fragments required for binding to the dataset. Specifically, we include $150$ generic inactives, $150$ inactives with the first two fragments, $150$ inactives with the first and third fragments, and $150$ inactives with the second two fragments. 

\begin{figure}[!ht]
    \centering
    \includegraphics[width=0.4\textwidth]{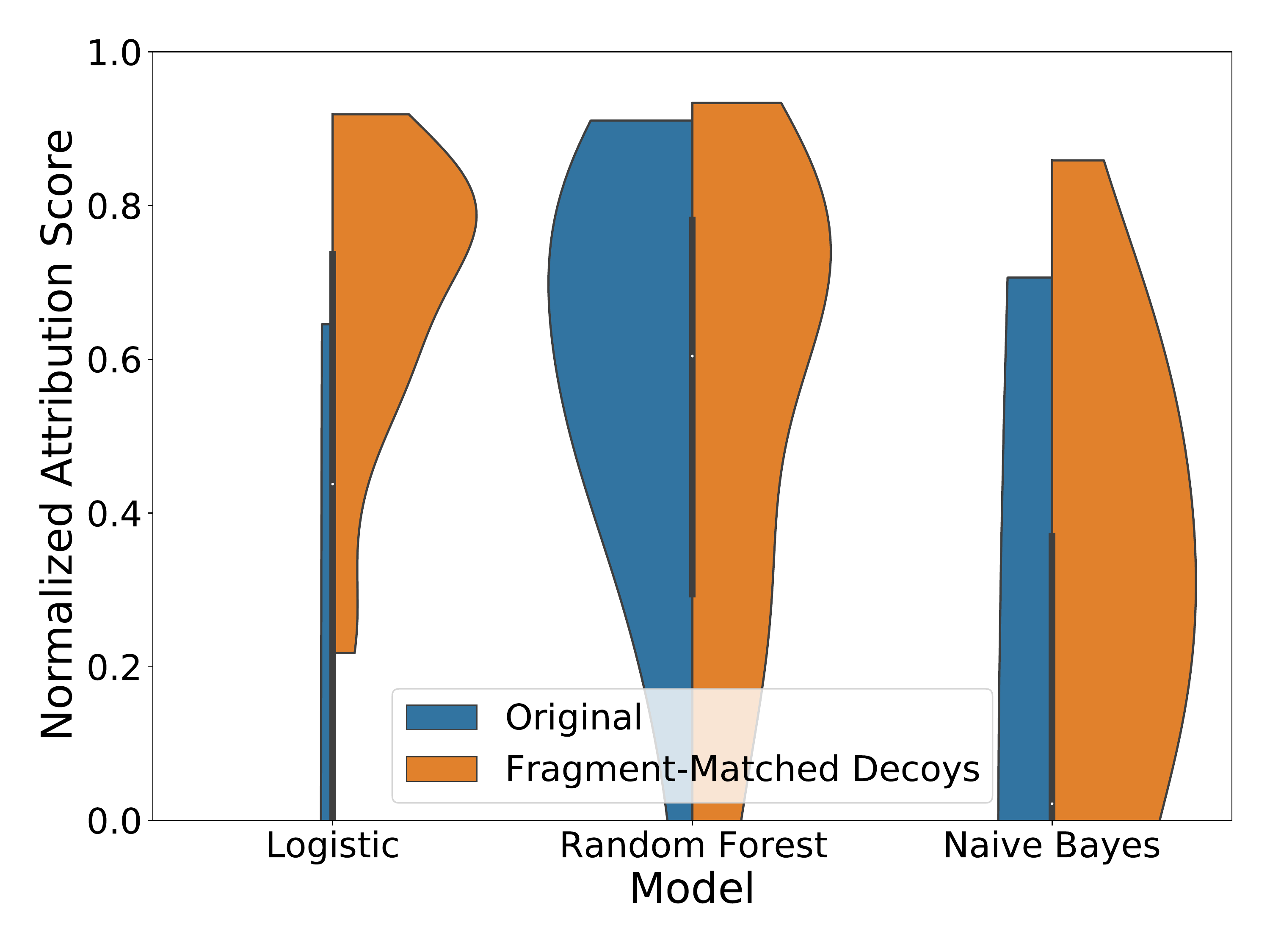}
    \caption{Effect of Adding Fragment-Matched Decoys on Normalized Attribution Score for Dataset 1. Adding fragment-matched decoys improves overall attribution scores, but they are still not perfect, at least partly due to high rates of false positives.}
    \label{fig:ToyDataDebiasEffect}
\end{figure}

We find that adding these fragment-matched decoys decreases the number of attribution false negatives (data not shown), and improves normalized attribution scores (Figure \ref{fig:ToyDataDebiasEffect}). However, our models are still not perfect, as they have plenty of attribution false positives and it is still possible to generate adversarial examples (data not shown). Thus adding fragment-matched decoys is necessary but not sufficient to improve the overall attribution accuracy of our models. When working with real world datasets, one could identify fragment-matched decoys by screening molecules with some, but not all, of the fragments observed in known actives. For example, combinatorial libraries where molecules are generated by linking fragments selected from fragment libraries would serve this purpose.

\subsection{Misattribution: Background Correlations}

To understand attribution false positives, we note that the features most correlated with activity in dataset 1 include a number of ethers that connect alkynes to benzenes or to alcohols. The alkynes, benzenes, and alcohols are part of the binding logic, but the ether is not. Similarly, in Figure \ref{fig:ToyDataThreeElemAdversarial}, we see that ethers are incorrectly included in the attributed fragments. This suggests that at least some attribution false positives are due to features that are highly correlated with activity but are not part of the binding logic. This could be caused by the fact that the features are not truly independent, i.e. some features may co-occur in dataset molecules more often than would be expected at random. To measure this, we compute the Pearson correlation between every pair of features across all molecules in dataset 1. This analysis reveals that ether is highly correlated with both the benzene and alkyne groups that are present in the binding logic, explaining why ether is highly correlated with activity. Since the dataset molecules are drawn randomly from ZINC12, these high inter-feature correlations can only be a result of spurious background correlations already present in the ZINC12 dataset. 

\begin{figure}[!ht]
    \centering
    \includegraphics[width=0.5\textwidth]{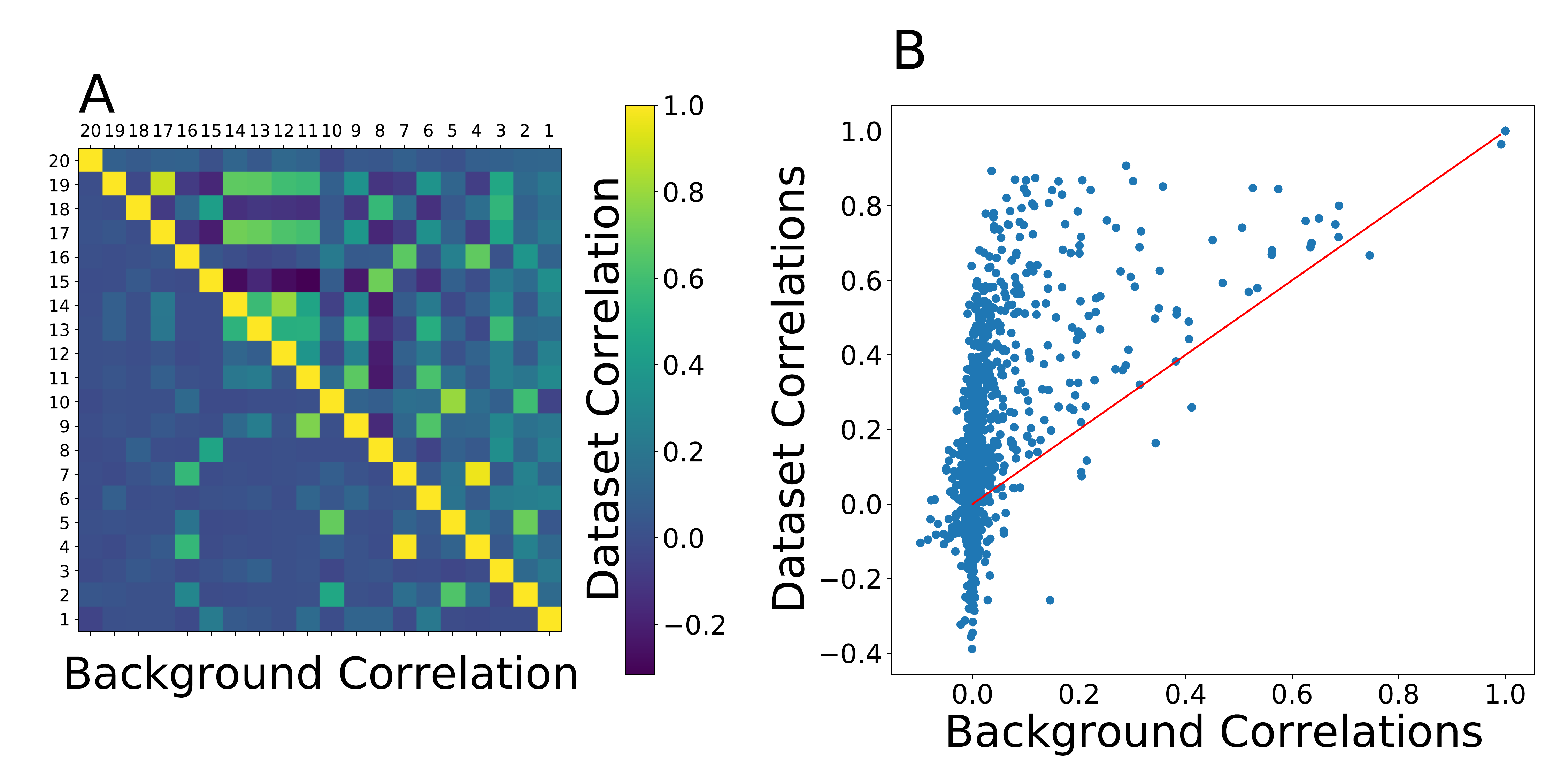}
    \caption{Correlation between Background and Dataset 1. (A) Top right: Heatmap of dataset 1 correlation matrix using hashed fingerprints. Bottom left: Background dataset correlation matrix from unhashed fingerprints. (B) Scatter plot of these correlations. We observe that many spurious correlations in dataset 1 are likely caused by the background. Correlations that are only present in the dataset are useful for determining the correct binding logic for each dataset.}
    \label{fig:ToyDatasetCorrelationMatrix}
\end{figure}

It has previously been shown that background correlations of hashed fingerprints are approximately drawn from the standard Marchenko-Pastur distribution that would be expected for random vectors drawn from a multivariate Gaussian distribution \cite{Lee2016}. This suggests that there is no additional correlation structure inherent in molecular fingerprint data. Despite this result, both the nature of small molecule structures and the manner in which fingerprints are constructed suggest that there should be correlations between bits that correspond to e.g. fragments and their substructures. To probe this more deeply, we computed the pair correlation matrix for a random sample of $4000000$ molecules from ZINC12 using unhashed ECFP6 fingerprints and found significant background correlation, as shown in Figure \ref{fig:ToyDatasetCorrelationMatrix}. This suggests that the hashing process obscures important information about the background correlation structure.

Our results indicate that some of the highest correlations in the hashed fingerprints observed in dataset 1 are related to this background correlation structure of the unhashed fingerprints, as shown by the cloud of points near the $y = x$ line in the scatter plot in Figure \ref{fig:ToyDatasetCorrelationMatrix}. Other correlations appear only in the dataset and are not present in the background; this generates the cloud of points surrounding the $y$-axis in the scatterplots of Figure \ref{fig:ToyDatasetCorrelationMatrix}. Our trained models need to distinguish informative correlations caused by the binding logic from spurious background correlations caused by molecular descriptors.

Successfully distinguishing between spurious and legitimate correlations does help uncover the correlations that correspond directly to the binding logic. After normalizing the dataset correlations by the background correlation, the highest correlations are between an alkyne and hydroxyl, a hydroxyl and a benzene, and an alkyne and a benzene: exactly the groups in the binding logic. Thus correlations that occur most strongly above background do correspond to the binding logic used to build the dataset.

\subsection{Comparison to Real Datasets}

In order to verify that attribution results for our \textit{in silico} datasets reflect those for real datasets, we examined comparative attribution scores. Real activity data was acquired from ChEMBL24.1 \cite{Davies2015, Gaulton2017} for a number of protein targets, with inactive molecules acquired from PubChem \cite{Mervin2018, Kim2016} following the procedure in \citet{Sundar2019}. We see similar degrees of disagreement between the models on the \textit{in silico} datasets and on the real datasets, as measured by comparative attribution scores (data not shown). We note that if two high-performing models like logistic regression and random forest disagree on attributions for specific molecules for which they make accurate binding predictions, then at least one must be wrong.

\section{Conclusions}

Our results suggest a number of important cautionary notes about the success of fingerprint-based protein/ligand binding models. Even high-performing models may misattribute, and using \textit{in silico} datasets to explicitly test model performance can help identify models that perform attribution correctly. We demonstrated that false negative attributions can be mitigated by adding fragment-matched decoys, where future work will test the sensitivity to accurate decoy selection. A corresponding approach for real datasets would be to screen molecules with a subset of the fragments present in known actives. Further, we have shown that the strongest false positives originate from correlations present in the background data that cause various fragments to be spuriously correlated with activity. Thus one key to mitigating false positives is to develop models that account for the background correlation structure.

\nocite{Unterthiner2014, Ramsundar2015, Duvenaud2015, Wallach2015, Kearnes2016}
\nocite{Verdonk2004, Cleves2008, Rohrer2009, Sheridan2013, Wallach2018, Liu2019, Sundar2019}
\bibliographystyle{icml2020}
\bibliography{Paper-Attribution}

\end{document}